%% file: main.tex
\documentclass[copyright,creativecommons,sharealike,noncommercial]{eptcs}
\providecommand{\event}{Gabriel Ciobanu (Ed.): Membrane Computing and\\Biologically Inspired Process Calculi 2009} %
\providecommand{\volume}{15}   %
\providecommand{\anno}{2009}   %
\providecommand{\firstpage}{1} %
\providecommand{\eid}{1}       %

\usepackage{breakurl}

\usepackage{amsmath}
\usepackage{amsthm}
\usepackage{amssymb} %

\usepackage{verbatim}
\usepackage{mimsymbols}

\usepackage{multicol}

\include{macros/generic}

\include{macros/mimcalc}
\include{macros/thenvs}

\title{A Process Calculus for Molecular Interaction Maps}
\author{Roberto Barbuti
  \institute{Dipartimento di Informatica\\Universit\`a di Pisa, Italy}
  \email{barbuti@di.unipi.it}
\and
  Andrea Maggiolo--Schettini
  \institute{Dipartimento di Informatica\\Universit\`a di Pisa, Italy}
  \email{maggiolo@di.unipi.it}
\and
  Paolo Milazzo
  \institute{Dipartimento di Informatica\\Universit\`a di Pisa, Italy}
  \email{milazzo@di.unipi.it}
\and
Giovanni Pardini
  \institute{Dipartimento di Informatica\\Universit\`a di Pisa, Italy}
  \email{pardinig@di.unipi.it}
\and
 Aureliano Rama
  \institute{Dipartimento di Informatica\\Universit\`a di Pisa, Italy}
  \email{rama@di.unipi.it}
}
\def\titlerunning{A Process Calculus for Molecular Interaction Maps}
\def\authorrunning{Barbuti, Maggiolo--Schettini, Milazzo, Pardini, Rama}

\begin{document}
\maketitle

\begin{abstract}
We present the \emph{MIM calculus}, a modeling formalism with a strong
biological basis, which provides biologically-meaningful operators for
representing the interaction capabilities of molecular species. The
operators of the calculus are inspired by the reaction symbols used in Molecular
Interaction Maps (MIMs), a diagrammatic notation used by biologists.
Models of the calculus can be easily derived from MIM diagrams, for which an
unambiguous and executable interpretation is thus obtained.
We give a formal definition of the syntax and semantics of the MIM
calculus, and we study properties of the formalism. A case study is also presented to show the use of the calculus for modeling biomolecular networks. 
\end{abstract}

\section{Introduction}
The use of formal methods in Systems Biology provides important advantages in the description and analysis of biological systems, since the structure and behavior of biological systems can be described unambiguosly and different analysis techniques can be applied to their study.
In this field, the most influential approach has been proposed by Regev, Shapiro and others in \cite{RSS00,RSS01,PRSS01,R02}, where the $\pi$--calculus process algebra \cite{M99,P95} is used to formalize biomolecular processes. Afterwards, many other formalisms originally developed by computer scientists to model systems of interacting components have been applied to Biology \cite{RS02,CDPB04,PRSS01,VG08}, and extended to allow more precise descriptions of the biological behaviors.
Other formalisms have also been developed expressly for being used in Biology \cite{BMMT06,BMM07,BMMTT07,Car05,CCDFS04,CD03,CH08a,LT06,DL04,PQ05,RPSCS04,BDP07,K03}.

Biologists have introduced graphical languages for describing bioregulatory networks. As an example, we quote \emph{Molecular Interaction Maps} (MIM) \cite{KAWP05}.
MIM diagrams are composed of nodes, representing molecular species, and edges connecting nodes, which represent the possible reactions among species. Edges can express different kinds of reactions, according to the used reaction symbol.
In this paper, we present a formalism which can be used for modeling and analyzing biological processes, called  \emph{MIM Calculus} (MIMc), which focusses on modeling the interaction capabilities of the involved elements.
MIMc is defined in the style of \emph{process calculi}, where each molecule appearing in the system is described by a term. However, unlike most of the previously proposed calculi for describing biological processes, which model reactions by means of process communication, MIMc provides high-level operators with a direct biological meaning.
For example, there are operators for expressing the creation of a bond between two compounds (such as a \emph{complexation}), and other biologically interesting events. 

The calculus has a strong relationship with Molecular Interaction Maps. 
The presented approach has a twofold advantage.
On one side, we can exploit the features of process calculi such as incremental definition of models, techniques for analysis and verification of properties, and easy development of simulators. On the other side, the correspondence of the operators of the calculus with biological interactions allows an immediate translation of Molecular Interaction Maps into MIMc.
Less immediate translations of Molecular Interaction Maps into more general formalisms can be
found in \cite{BLMMPR09,BFP07,CPQ05}.
Remark that the aim of the paper is to propose a calculus whose operators have a direct correspondence with the ones of Molecular Interaction Maps. Thus all the main MIM operators are considered, without any investigation about a minimal set of them able to encode all the others.

The paper is structured as follows. After recalling the Molecular Interaction Maps, in section 2 we introduce the MIM calculus. In section 3 we study the relationship between the MIM calculus and Molecular Interaction Maps, and we establish conditions under which a term of the calculus is a formal representation of a MIM diagram. In section 4 we show an example of modeling with the MIM calculus, and in section 5 we draw some conclusions.

\subsection{Molecular Interaction Maps}
MIM diagrams provide a static view of the molecular species in a system, and their possible interactions. Interactions are represented by lines connecting nodes representing species, and the meaning of each interaction depends on the symbol used to draw the line. Each molecular species can appear only once. Moreover, since the diagram is static, it does not contain any information about number of molecules (concentration) of the molecular species.

\begin{figure}[htp]
 \centering
 \includegraphics[scale=1]{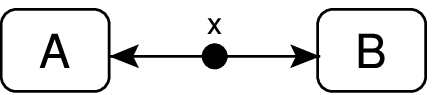}
 \caption{An example of MIM diagram.}
 \label{fig:mim-speciesexample}
\end{figure}

\begin{figure}[t]
\centering
\begin{minipage}[c]{0.31\linewidth}
\centering
\includegraphics[width=4.5cm]{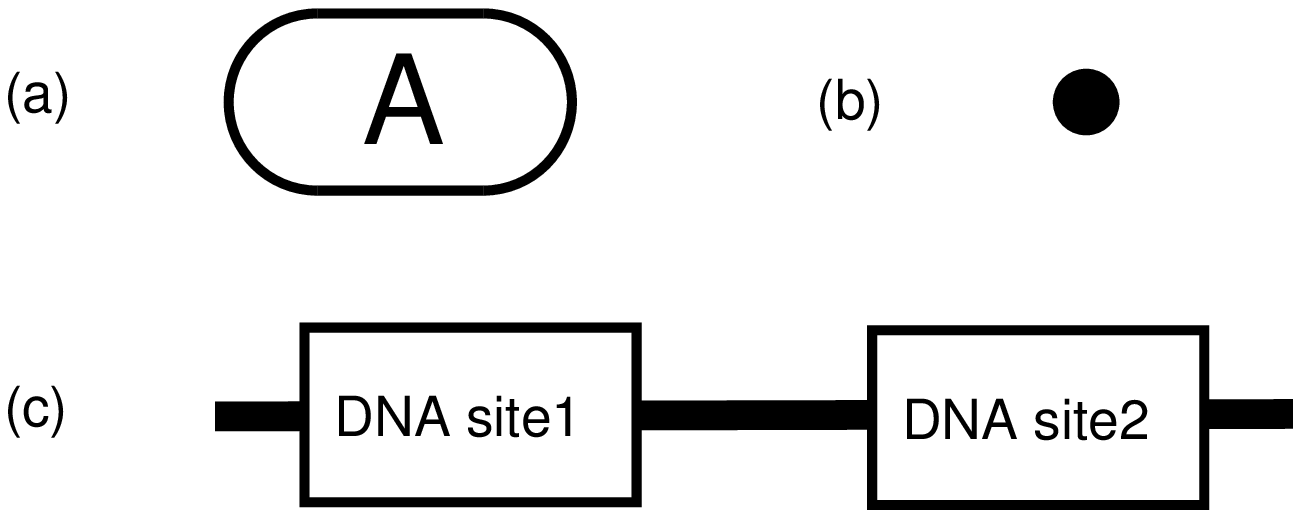}
\caption{Species in MIMs.}
\label{fig:mim-species}
\end{minipage}
\begin{minipage}[c]{0.31\linewidth}
\centering
\includegraphics[width=4.5cm]{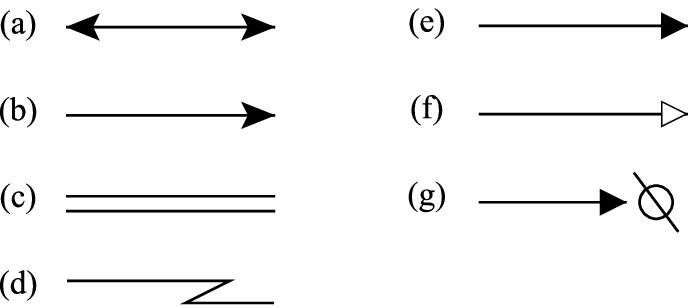}
\caption{Reaction symbols.}
\label{fig:mim-symbols}
\end{minipage}
\begin{minipage}[c]{0.34\linewidth}
\centering\includegraphics[width=4.5cm]{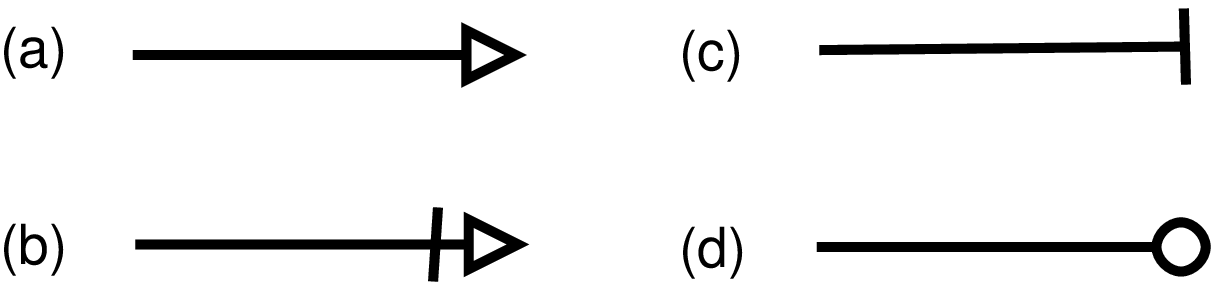}
\caption{Contingency symbols.}
\label{fig:mim-contingencysymbols}
\end{minipage}
\end{figure}

Three classes of molecular species can be represented: \emph{elementary species} (fig. \ref{fig:mim-species}a), \emph{complex species} (fig. \ref{fig:mim-species}b) and \emph{DNA sites} (fig. \ref{fig:mim-species}c).
Complex species represent either a combination of elementary species or a modified elementary species.
Figure \ref{fig:mim-speciesexample} shows a simple MIM diagram, containing the elementary species $A$ and $B$ which can interact.
A named elementary species is drawn as a rounded box, containing its name. A complex molecular species, resulting from an interaction, is depicted as a bullet on the corresponding interaction line. For instance, in Figure \ref{fig:mim-speciesexample}, the complex species obtained by the binding of $A$ and $B$ is represented by the node $x$ on the interaction line.

MIM diagrams allow representing two kinds of interactions: \emph{reactions}, which act on molecular species, and \emph{contingencies}, which act on reactions or other contingencies. An interaction symbol represents a possible interaction that can happen if certain state conditions hold. Interactions can have a kinetic constant $k$ associated with them, that is used to model its ``occurrence'' rate. Conceptually, a higher kinetic constant means that the interaction is more likely to happen than an interaction with a lower kinetic constant.

For defining MIM calculus we consider the reaction symbols shown in Figure \ref{fig:mim-symbols}. Note that they are only a subset of the reaction symbols available for use in a MIM diagram.

\begin{itemize}
\item \emph{Non-covalent binding} (Figure \ref{fig:mim-symbols}a): denotes the reversible binding of the two pointed species: a molecule of the first species can bind to a molecule of the second species, forming a compound. Two species joined in a non-covalent binding can eventually dissociate again.
\item \emph{Covalent modification} (Figure \ref{fig:mim-symbols}b): denotes the covalent modification of the pointed species; the modification type (such as phosphorylation or acetylation) is written at the tail.
\item \emph{Covalent binding} (Figure \ref{fig:mim-symbols}c): denotes a covalent bond of the two connected species.
\item  \emph{Cleavage of a covalent bond} (Figure \ref{fig:mim-symbols}d): denotes the possibility of a covalent bond at the head (right end) to be broken by the presence of the species at the tail (left end). This symbol points from a species to a reaction symbol  representing covalent binding.
\item  \emph{Stoichiometric conversion} (Figure \ref{fig:mim-symbols}e): denotes the conversion of the species at the tail of the arrow, called reactant, into a corresponding number of product species, i.e. the species written at the tail of the arrow disappears, while the pointed ones appear.
\item  \emph{Lossless production} (Figure \ref{fig:mim-symbols}f):  it is similar to the stoichiometric conversion, but without loss of the reacting species.
\item  \emph{Degradation} (Figure \ref{fig:mim-symbols}g): means that molecules of the species can disappear.
\end{itemize}

The following contingency symbols, shown in Figure \ref{fig:mim-contingencysymbols}, are provided by MIM diagrams:
\begin{itemize}
 \item \emph{Stimulation} (Figure \ref{fig:mim-contingencysymbols}a): means that the molecule of the species at left end stimulates the pointed reaction;
 \item \emph{Requirement} (Figure \ref{fig:mim-contingencysymbols}b): means that the molecule of the species at left end is required in order for the pointed reaction to happen;
\item \emph{Inhibition} (Figure \ref{fig:mim-contingencysymbols}c): the presence of the species at the tail (left end) inhibits the possibility for the pointed interaction to happen;
\item \emph{Catalysis}  (Figure \ref{fig:mim-contingencysymbols}d): means that the pointed reaction have a much higher reaction rate if the species is present than if it is not.
\end{itemize}

\begin{figure}[htp]
 \centering
 \includegraphics[scale=0.2]{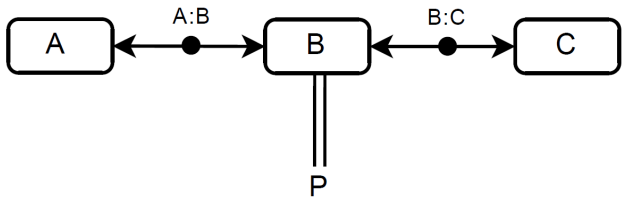}
 \caption{An example of MIM diagram.}
 \label{fig:mim-interpretations}
\end{figure}

\paragraph{Interpretation of MIM diagrams}
There are three different interpretations for MIM diagrams \cite{KAWP05,KAKWP06}: \emph{explicit}, \emph{combinatorial} and \emph{heuristic}. Each interpretation is suited to a different purpose, depending on the application. %
They differ in how interactions between indirectly connected species are considered. Figure \ref{fig:mim-interpretations} shows a small example of MIM, which explicitly shows the bindings between A and B, yielding A:B; the binding between B and C, yielding B:C; and the possible phosphorylation of B, yielding pB. Some questions arise, such as if a complex between A:B and C, yielding  (A:B):C, can form. Or, similarly, if pB and A can bind.
The three different interpretations address this issue, by stating which interactions are possible.

A MIM diagram, in its explicit interpretation, depicts each possible reaction: an interaction line applies only to the molecular species directly connected to it. In this interpretation the order of bindings can be easily extracted from the diagram. In Figure \ref{fig:mim-interpretations}, the explicit interpretation only allows the formation of A:B, B:C and pB. Explicit maps can be built using only a subset of MIM symbols: all contingencies symbols may, on the whole, be represented by a set of reaction symbols (\cite{KAWP05}).
Explicit maps without contingencies can be readily used for computer simulation.

Besides the complexes that are allowed by the explicit interpretation, in the combinatorial interpretation of MIM diagrams, an interaction line represents an implicit set of complexes and, hence, of reactions. In particular, each interaction line represents all those reactions between the interacting species, in each possible combination of their binding and modification states. 
In Figure \ref{fig:mim-interpretations}, the combinatorial interpretation always allows A and B to bind, regardless of the fact that B is free, bound to C, or phosphorylated. This ``transitivity'' means that an interaction symbol applies indirectly to species through other interaction symbols. A main advantage of the combinatorial interpretation is this ability to synthesize with a few symbols a
large number of possible complexes and reactions, making MIM a compact notation.

Finally, like the combinatorial interpretation, the heuristic one allows all the complexes that are permitted by the explicit interpretation, with the difference that it does not specify whether each of the combinatorial possibilities may or may not occur, either because of lack of knowledge or because some contingency symbols have been omitted to avoid overcrowding the diagram. Thus, heuristic MIM diagrams are used to depict only what is known, leaving unspecified what still has to be discovered.
\section{MIM Calculus}
In this section we formally introduce the syntax and semantics of the MIM calculus.
MIM calculus is defined in the style of process calculi, where an agent represents a molecule of a certain named species.
Names $A,B,C,\ldots$ are used to identify the different elementary species, and we denote by $\calE$ the set of names of elementary species.
We also assume a set $\calE_c$ whose elements denote types of covalent modifications (such as phosphorylations).

\begin{definition}[Syntax] \emph{Processes} $P$, \emph{named species} $S$ and capabilities $\mu$ of the \emph{MIM calculus} are defined by the following grammar:
\begin{align*}
P	&\IS \nil \OR S \OR P \mid P  
		&\qquad  \gamma &\IS (\nu, \iota) \complexation{N} \mu	&&\quad\textit{(non-covalent binding)} \\
S	&\IS \mu . IS
		&\qquad &\OR (\nu, \iota) \covalentbind{N} \mu		&&\quad\textit{(covalent binding)} \\
IS	&\IS A \OR S:S \OR \bond{qS} \OR \bond{S S}
		&\qquad &\OR (\nu, \iota) \covalentmod{q} \mu		&&\quad\textit{(covalent modification)} \\
\mu	&\IS rec\ x.\mu \OR M \OR x
		&\qquad &\OR (\nu, \iota) \cleavage{N}			&&\quad\textit{(cleavage)} \\
M	&\IS \varnothing  \OR M + M  \OR  \gamma
		&\qquad &\OR (\nu, \iota) \conversion 		P &&\quad\textit{(conversion)} \\
	&
		&\qquad &\OR (\nu, \iota) \losslessprod 		P &&\quad\textit{(lossless production)}
\end{align*}
where $\nil$ is the empty process,  $A \in \calE$ denotes an elementary species name, $q \in \calE_c$ denotes the type of modification, $x \in \calX$ is a variable, and $N, \nu, \iota$ denote species names, which are elements of the set $\calN$ of terms $S$ without capabilities.
For the sake of legibility we shall often use round brackets and we shall systematically enclose capabilities in curly brackets.
\end{definition}

Terms $P$ of the calculus are made of a composition of molecules $S$, by means of the parallel operator $\_ \pipe \_$.
Each molecule is of the form $\mu.IS$, where $IS$ describes the structure of the molecule, and $\mu$ describes its interaction capabilities.
In particular, $IS$ denotes either an elementary molecule of species $A$, or a compound molecule.
In the case of compound molecules, $IS$  is made of the single molecules forming the compound, combined by means of different syntactical operators specifying the kind of bond that keeps the molecules together: a \emph{non-covalent bond} $S_1 : S_2$ between the species $S_1$ and $S_2$, a \emph{covalent modification} $\bond{q S}$ of species $S$, or a \emph{covalent bond} $\bond{S_1 S_2}$ between $S_1$ and $S_2$.
Note that  the capabilities of each molecule forming a compound are retained in the compound description.

For example, term $\{\gamma\}.A$ models a molecule of species $A$, having a single interaction capability $\gamma$. A complex formed of two simple molecules $A$ and $B$ can instead be represented as $\mu_1.(\mu_2.A : \mu_3.B)$, where $\mu_1$ are the capabilities of the compound, and $\mu_2,\mu_3$ are the capabilities of molecules $A$ and $B$, respectively.

We denote the set of species by $\calS$, and identify $\calN \subset \calS$ as its subset of named species without capabilities, i.e. where each $\mu$ is empty ($\mu = \varnothing$). We assume a function $\inactive{\cdot} : \calS \to \calN$ that strips all the capabilities from a named species $S \in \calS$. For example, $\inactive{\mu_1.(\mu_2.A : \mu_3.B)} = \varnothing.(\varnothing.A : \varnothing.B)$. Moreover, we often avoid writing empty capabilities when no ambiguities arise, therefore we simply write  $A:B$ instead of $\varnothing.(\varnothing.A : \varnothing.B)$. This function is extended to processes $\inactive{\cdot} : \calP \to \PowerSet(\calN)$ as $\inactive{S_1 \pipe \cdots \pipe S_n} = \inactive{S_1} \cup \cdots \cup \inactive{S_n}$.

The calculus allows expressing different capabilities for molecules. Operator $\complexation{N} \mu$ means that a species can form a non-covalent bond with a molecule of species name $N$. The result will be a compound of the form $\_ : \_$ made of the two involved species, and with capabilities $\mu$. Similarly, operator $\covalentbind{N} \mu$ means that a species can form a covalent bond with a $N$, resulting in a compound of the form $\bond{S_1\ S_2}$ with capabilities $\mu$. The operator for covalent modification $\covalentmod{q} \mu$, similarly produces a compound $\bond{qS}$ with capabilities $\mu$.
The operator for cleavage $\cleavage{N}$ means that a molecule can break the covalent bond specified by $N$, where $N$ has to be either of the form of $\bond{N_1 N_2}$ or $\bond{q N}$. Finally, there are the operators $\conversion P$, for expressing a conversion of a molecule into other molecules, and $\losslessprod P$ for a lossless production of molecules. In both cases, the resulting molecules are represented by a process $P$.

We allow recursive definitions of capabilities, by means of the \emph{recursion} operator $rec$. As always, $rec\ x.\mu$ binds the free occurrences of the variable name $x$ in $\mu$. We assume a \emph{substitution} function $\mu [\mu' / x]$ for replacing each free occurrence of $x$ in $\mu$ with $\mu'$. The substitution function is also extended to processes. We use the notation $rec\ \tilde{x} . \mu$ with $\tilde{x} = x_1,\ldots,x_n \in Var^*$ as an abbreviation for $rec\ x_1. \cdots . rec\ x_n . \mu$.

Names $\nu, \iota$ are used to express contingencies on the application of an operator, depending on the species appearing in the environment. The former, $\nu$, expresses the species that \emph{must} be present \emph{(promoters)}, while the latter, $\iota$, expresses those that must be \emph{absent} \emph{(inhibitors)}. We omit writing contigencies when they are empty.

\begin{figure}[htp]
 \centering
 \includegraphics[scale=1]{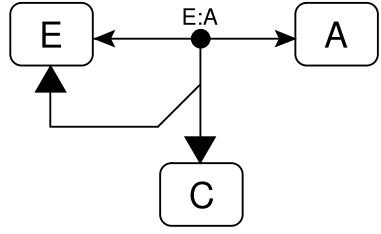}
 \caption{An example of recursive MIM}
 \label{fig:mim-recexample}
\end{figure}

To give an example of a term in which recursive capabilities are used, let us consider a system in which substrate $A$ is transformed into product $C$ by the enzyme $E$. The MIM diagram in Figure \ref{fig:mim-recexample} shows that enzyme $E$ binds to $A$ and the complex $E:A$ is subsequently transformed into $C$ and $E$, thus recreating the enzyme. The enzyme $E$ can be modeled in MIMc by the following term: $rec\ x. \{ \complexation{A} \{ \conversion ( x.E \pipe \varnothing.C ) \} \}.E$.

\begin{definition}[Structural congruence]\label{def:structcongr}
The congruence relations $\equiv_x$ on the syntactical categories $x \in \{P,S,IS,\mu,M,\gamma\}$ of the calculus  
are the least equivalence relations closed under syntactical operators and such that the following laws hold:
\begin{enumerate}
 \item $P_1 \pipe P_2 \equiv_P P_2 \pipe P_1$,\quad $P_1 \pipe (P_2 \pipe P_3) \equiv_P (P_1 \pipe P_2) \pipe P_3$,\quad $P \pipe \nil \equiv_P P$;
 \item $S_1 : S_2 \equiv_{IS} S_2 : S_1$,\quad $\bond{S_1 S_2} \equiv_{IS} \bond{S_2 S_1}$;
 \item $M_1 + M_2 \equiv_M M_2 + M_1$, $M_1 + (M_2 + M_3) \equiv_M (M_1 + M_2) + M_3$, $M + \varnothing \equiv_M M$, $M + M \equiv_M M$;
 \item ($\alpha$-conversion) $\mu_1 \equiv_\mu \mu_2$ if they differ only on bound names;
 \item $rec\ x.\mu \equiv_\mu \mu[rec\ x.\mu / x]$.
\end{enumerate}
\end{definition}
\noindent We omit the indication of $x$ in $\equiv_x$ when no ambiguities arise.

We propose now a reduction semantics for the MIM calculus, given in terms of a Labelled Transition System (LTS) representing the possible evolutions of a term. The labels of the LTS are actions identifying the reactions that each single transition models. All the possible actions $Act$ are of the following forms: 
 (a) $N_1 \vcomplexation N_2$,
 (b) $N_1 \vdecomplexation N_2$,
 (c) $N \vconversion \{N_1,\ldots,N_k\}$,
 (d) $N \vlosslessprod \{N_1,\ldots,N_k\}$,
 (e) $N_1 \vcovalentbind N_2$,
 (f) $N \vcleavage \bond{N_1 N_2}$,
 (g) $q \vcovalentmod N$,
 (h) $N \vcleavage \bond{q N_1}$,
which respectively represent
(a) the creation of a non-covalent bond, 
(b) the cleavage of a non-covalent bond,
(c) a conversion,
(d) a lossless production,
(e) the creation of a covalent bond,
(f) the cleavage of a covalent bond,
(g) a covalent modification,
(h) the removal of a covalent modification.

\begin{definition}[Reduction semantics] The \emph{reduction semantics of MIM calculus} is the relation $\ltrans{\alpha}$ on processes such that
\begin{align}
 P \ltrans{\alpha} P' \quad \text{iff} \quad \exists \iota \in \calN.\; P \ltrans{(\emptyset, \iota)\ \alpha} P'    \label{eq:sem0}
\end{align}
where $\alpha \in Act$ is an action that represents the capability of $P$ used for the reduction step, and $\ltrans{(\nu, \iota) \alpha}$, with $\nu,\iota \in \calN$, is the least relation on processes, closed under structural congruence $\equiv_P$, and satifying the following inference rules:
{\allowdisplaybreaks
\begin{gather}
 \InfRule{\mu_1 = \{ X + (\nu, \iota) \complexation{\inactive{S_2}} \mu \}
		\AND \alpha = \inactive{S_1} \vcomplexation \inactive{S_2}
 		\AND  \inactive{S_1},\inactive{S_2} \not\in \iota }
		{ \mu_1.S_1 \mid \mu_2.S_2  \ltrans{(\nu, \iota)\ \alpha}   \mu.( \mu_1.S_1 : \mu_2.S_2) } \label{eq:sem-complx}  \NR
 \InfRule{ \alpha = \inactive{S_1} \vdecomplexation \inactive{S_2} }
		{\mu.(S_1:S_2) \ltrans{(\emptyset, \emptyset)\ \alpha} S_1 \pipe S_2} \label{eq:sem-decomplx} \NR
\InfRule{ \mu = \{ X + (\nu, \iota) \conversion P \}
 		\AND  \alpha = \inactive{S} \vconversion \inactive{P}
		\AND  \inactive{S} \not\in \iota }
		{\mu.S \ltrans{(\nu, \iota)\ \alpha} P } \label{eq:sem-conversion}\NR
\InfRule{ \mu = \{ X + (\nu, \iota) \losslessprod P \}
 		\AND  \alpha = \inactive{S} \vlosslessprod \inactive{P}
		\AND  \inactive{S} \not\in \iota }
		{\mu.S \ltrans{(\nu, \iota)\ \alpha} \mu.S \pipe P } \label{eq:sem-losslessprod}\NR
 \InfRule{\mu_1 = \{ X + (\nu, \iota) \covalentbind{\inactive{S_2}} \mu \}
		\AND  \alpha = \inactive{S_1} \vcovalentbind \inactive{S_2}
		\AND  \inactive{S_1},\inactive{S_2} \not\in \iota  }
		{ \mu_1.S_1 \pipe \mu_2.S_2  \ltrans{(\nu, \iota)\ \alpha}   \mu.\bond{(\mu_1.S_1)(\mu_2.S_2)} } \label{eq:sem-covalentbind} \NR
 \InfRule{ \mu = \{ X + (\nu, \iota) \cleavage{\inactive{\bond{S_1 S_2}}} \}
		\AND  \alpha = \inactive{S} \vcleavage \inactive{\bond{S_1 S_2}}
		\AND \inactive{S},\inactive{\bond{S_1 S_2}} \not\in \iota  }
		{\mu.S \pipe \mu'.\bond{S_1 S_2} \ltrans{(\nu, \iota)\ \alpha} \mu.S \pipe S_1 \pipe S_2} \label{eq:sem-cleavagebond} \NR
 \InfRule{\mu_1 = \{ X + (\nu, \iota) \covalentmod{q} \mu \}
		\AND  \alpha = q \vcovalentmod \inactive{S_1}
		\AND  \inactive{S_1}  \not\in \iota  }
		{ \mu_1.S_1  \ltrans{(\nu, \iota)\ \alpha}   \mu.\bond{q(\mu_1.S_1)} }  \label{eq:sem-covalentmod} \NR
 \InfRule{ \mu = \{ X + (\nu, \iota) \cleavage{\inactive{\bond{q S_1}}} \}
		\AND  \alpha = \inactive{S} \vcleavage \inactive{\bond{q S_1}}
		\AND  \inactive{S},\inactive{\bond{q S_1}} \not\in \iota  }
		{\mu.S \pipe \mu'.\bond{q S_1} \ltrans{(\nu, \iota)\ \alpha} \mu.S \pipe S_1} \label{eq:sem-cleavagemod} \NR
 \InfRule{ P \ltrans{(\nu, \iota)\ \alpha} P'  \AND   \inactive{Q} \cap \iota = \emptyset  \AND  \nu' = \nu \setminus \inactive{Q} }
		{ P \pipe Q \ltrans{(\nu', \iota)\ \alpha} P' \pipe Q } \label{eq:sem-parallel} %
\end{gather}
}%
\end{definition}

Rule \ref{eq:sem-complx} deals with the creation of a non-covalent bond between molecules $\mu_1.S_1$ and $\mu_2.S_2$, thus giving rise to a complex $\mu.( \mu_1.S_1 : \mu_2.S_2)$. Note that the rule requires that molecule $\mu_1.S_1$ has the capability of binding with a molecule with name $\inactive{S_2}$ (the symmetric capability is not required for $\mu_2.S_2$). Rule \ref{eq:sem-decomplx} deals with the cleavage of a non-covalent bond. There are no conditions for the cleavage. Rule \ref{eq:sem-conversion} deals with the conversion of a molecule $\mu.S$ into a number of other molecules when $\mu.S$ has the proper capability. Rule \ref{eq:sem-losslessprod} deals with the lossless production, namely with the case in which $\mu.S$ produces a number of molecules without disappearing. Rules \ref{eq:sem-covalentbind} and \ref{eq:sem-cleavagebond} are the analogs of rules \ref{eq:sem-complx} and \ref{eq:sem-decomplx} for the case of covalent binding. The unbinding, expressed by rule \ref{eq:sem-cleavagebond}, requires the presence of a molecule 
$\mu.S$ having the capability of breaking the bond. Rules \ref{eq:sem-covalentmod} and \ref{eq:sem-cleavagemod} deal with molecule covalent modification of a type $q$ and with the removal of the modification, respectively.

All the rules are applicable only if the molecules involved are not inhibitors of the transformation itself. 
Rule \ref{eq:sem-parallel} is used to apply a step of the reduction to the parallel composition of processes.
The conditions of the rule ensure that the step is not forbidden by any of the molecules present in the composition.
Moreover, these conditions, together with the fact that we need that the set $\nu$ of promoters is empty to actually do the reduction step (defined in \ref{eq:sem0}), ensure that  all the promoters of the capability used for that step are present in the parallel composition of processes.
As usual, we define $\ltrans{}^*$ as the reflexive and transitive closure of relation $\ltrans{\alpha}$.

We now define \emph{contexts}, which represent terms with a hole, denoted as $\hole$. 
The hole corresponds to the collection of capabilities of a molecule, and therefore the hole occurs in the position of a capability $\mu$.
Conceptually, contexts allow identifying the position inside a term in which a molecule of a certain species appears. 
\begin{definition}[Context] Contexts of MIM calculus are defined by the following grammar:
\begin{align*}
C	&\IS P \pipe S_c
		&\qquad \gamma_c &\IS (\nu, \iota) \complexation{N} \mu_c	&&\quad\textit{(non-covalent binding)} \\
S_c	&\IS \mu_c . IS \OR \mu.IS_c
		&\qquad &\OR (\nu, \iota) \covalentbind{N} \mu_c		&&\quad\textit{(covalent binding)} \\
IS_c	&\IS S:S_c \OR \bond{q S_c} \OR \bond{S S_c}
		&\qquad &\OR (\nu, \iota) \covalentmod{q} \mu_c		&&\quad\textit{(covalent modification)} \\
\mu_c	&\IS \hole  \OR M + \gamma_c
		&\qquad &\OR (\nu, \iota) \conversion 	C	 &&\quad\textit{(conversion)} \\
    &
		&\qquad &\OR (\nu, \iota) \losslessprod 	C	 &&\quad\textit{(lossless production)}
\end{align*}
\end{definition}
The syntax ensures that exactly one hole is present in a context. 
Given $S_c$, the hole can occur 
either in the capabilities of $S_c$ itself, when $S_c = \mu_c.IS$, or in the capabilities of one of the molecules of which the molecule is composed, $S_c = \mu.IS_c$.
Given a capability with a hole $\mu_c$, the hole can be either the capability itself $\hole$, or 
it can occur in one of the capabilities appearing in $\mu_c$.
In particular, if $\mu_c = \gamma_c$ (a basic capability) the hole can occur inside the capabilities of the species
$\gamma_c$ allows producing.
Given a context $C$, its hole can be substituted with a capability $\mu$, giving a process denoted $C[\mu]$.

For example, the context $C_1 = \mu_1.(\mu_2.A : \hole.B)$ represents a molecule complex $A:B$ in which the hole refers to the capability of $B$ forming the complex. Context $C_1$ can be applied to a capability $\mu_3$ obtaining $C_1[\mu_3] = \mu_1.(\mu_2.A : \mu_3.B)$. Note that in this case, the hole is relative to a species named $B$, and this is clearly visible from the syntax of the context. However, in other cases, the name of the species relative to a hole is not directly present in the syntax of the context. For example, the hole in context $C_2 = \{\complexation{B}\hole + \losslessprod P  \}.A$ 
is relative to the species obtained as a complexation between $A$ and $B$, whose name is $A:B$, which is not directly present in the syntax of the context.
In order to extract, from a given context, the name of the species relative to the hole, we use a function $name: \calC \to \calN$ defined as follows. 

\begin{definition} Function $name$, from contexts to molecular names $\calN$, and $name'(\mu_c, N)=N$, from a context $\mu_c$ and a name $\calN$ to a name $\calN$, are recursively defined as follows:
\begin{multicols}{2}
 \begin{align}
  name( P \pipe S_c) &= name(S_c)\\
  name( \mu . IS_c ) &= name( IS_c )\\
  name( S : S_c ) &= name( S_c )\\
  name( \bond{q S_c} ) &= name( S_c )\\
  name( \bond{S S_c} ) &= name( S_c )\\
  name( \mu_c . IS ) &= name'( \mu_c, \inactive{IS} )  \label{eq:contextname1}
 \end{align}
  \begin{align}
  name'( \hole, N ) &= N \\
  name'( (\nu,\iota) \complexation{N'} \mu_c ,N) &= name'( \mu_c, N:N' )\\
  name'( (\nu,\iota) \covalentbind{N'} \mu_c ,N) &= name'( \mu_c, \bond{N\,N'} )\\
  name'( (\nu,\iota) \covalentmod{q} \mu_c ,N) &= name'( \mu_c, \bond{qN} )\\
  name'( (\nu,\iota) \conversion C ,N) &= name( C )\\
  name'( (\nu,\iota) \losslessprod C ,N) &= name( C )\\
  name'( M + \gamma_c ,N) &= name( \gamma_c ,N)
 \end{align}
\end{multicols}
\end{definition}
Definition of function $name$ is given by two mutually recursive functions $name$ and $name'$. In particular, $name'$ takes two parameters, a capability context $\mu_c$ and a name $N$, where $N$ is the name of the species with which this capability is associated. The function $name'$ is used in equation \ref{eq:contextname1}, where extracting the name, relative to the hole, from a context $\mu_c . IS$ is reduced to extracting the name from $\mu_c$, knowing that the capability $\mu_c$ is relative to a species named $\inactive{IS}$.

\paragraph*{Example}

\begin{figure}[htp]
 \centering
 \includegraphics[scale=1]{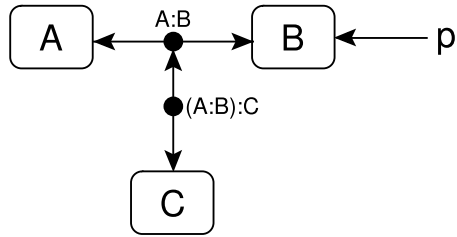}
 \caption{A MIM diagram.}
 \label{fig:mim-example}
\end{figure}

Consider an example of a MIM process which represents a molecular system described by the MIM diagram shown in Figure \ref{fig:mim-example}.

A MIM process, differently from a MIM diagram, represents both the possible interactions among the species and the number of molecules that are present in the system.
The following MIM process corresponds to a system with species $A$, $B$ and $C$, with the interaction capabilities described by the diagram in Figure \ref{fig:mim-example}, and in which there are two molecules of $A$, two of $B$ and one of $C$:
\newcommand{\oo}{(\emptyset,\emptyset)}
\begin{align*}
 P = &\{ \complexation{B} \{ \complexation{c}  \varnothing \}  \} . A  \pipe \\
	&\{ \complexation{B} \{ \complexation{c}  \varnothing \}  \} . A \pipe \\
	&\{ \complexation{A} \{ \complexation{c}  \varnothing \}  +  \covalentmod{p} \varnothing \} . B \pipe \\
	&\{ \complexation{A} \{ \complexation{c}  \varnothing \}  + \covalentmod{p} \varnothing \} . B \pipe \\
	&\{ \complexation{A:B} \varnothing \} . C
\end{align*}
In the process $P$, the species $A$ can complex with $B$, producing a molecule able to complex with $C$. Species $B$ is able to perform the same reaction of $A$, and it can also be phosphorilated. Finally, species $C$ can complex with $A:B$.
Recall that the non-covalent bond of $A:B$ and $C:(A:B)$ can dissociate autonomously as described by rule \ref{eq:sem-decomplx} of the reduction semantics.

\section{Consistency}
In this section we investigate the relationship between Molecular Interaction Maps and MIM Calculus.
We propose some \emph{consistency} definitions, with the aim of identifying the terms of the calculus which could be formal representation of a MIM diagram.
The first difference between MIM diagrams and the MIM calculus is that diagrams provide a static view of the species of a system, and of the interactions which can occur among the species, while MIM calculus allows representing single molecules and provides a semantics for deriving the evolution of the described system.
MIM diagrams are also restricted with regards to the capabilities of species, since each single molecular species can appear only once in a diagram.
For this reason, the informal interpretation of MIM diagrams assumes that the capabilities of each molecule depend only on the species of the molecule. Capabilites are irrespective, for example, of the different reactions that might produce a molecule of that species.
On the contrary, the MIM calculus allows representing single molecules $\mu.IS$, and different molecules of the same species might have different capabilities. For example, term $P = \{\complexation{B}\mu_1\}.A \pipe \varnothing.A \pipe P'$ contains two molecules of species $A$ with different capabilities: the first one can bind to another molecule of species $B$, while the second one has no capabilities. Note that these two molecules of species $A$ could have been obtained as a result of other reactions (for example, by transformation of other molecules), hence during any evolution of the system there may be some states in which all the molecules of the same species have the same capabilities while, in other states, this is not true.
It appears to be of particular interest to establish which terms of MIMc represent MIM diagrams, in the sense that a MIM diagram can be associated with a term, and the term evolves in accordance with the behavior intended by the diagram.
One may also ask that in a term molecules of a certain species always have the same capabilities. This captures the constraint of uniqueness of species in MIM diagrams.

For this purpose, we present three different definitions of consistency of MIM calculus terms, namely \emph{semantic consistency}, \emph{(weak) syntactic consistency}, and  \emph{strong syntactic consistency}.
Semantic consistency is the weakest form of consistency, and takes into account only terms that can be reached from the initial state. This form of consistency requires that, whenever a molecule of a certain species named $N$ is produced (i.e. a molecule $S$, with $\inactive{S} = N$, appears in the top-level parallel composition), it always has the same capabilities.
\begin{definition}[Semantic Consistency] A term $P$ is \emph{semantically consistent} iff
 \begin{equation*}
	\forall \mu_1.S_1, \mu_2.S_2, P', P''.\quad \inactive{S_1} \equiv \inactive{S_2}  \text{ and } P \ltrans{}^{*} \mu_1.S_1 \pipe P'  \text{ and } P \ltrans{}^{*} \mu_2.S_2 \pipe P'' \text{ implies }  \mu_1 \equiv \mu_2
 \end{equation*}
\end{definition}

The following definitions of syntactic consistencies deal instead with the species that syntactically appear in a term. The weak form requires that the capabilities of each molecule of a species appearing in the term, including those forming compound molecules and those that can be obtained as the result of reactions, always have the same capabilities. The definitions make use of contexts, in order to precisely identify the position inside the term in which molecules of a species (with their capabilities) appear.

\begin{definition}[(Weak) Syntactic Consistency] A term $P$ is \emph{(weakly) syntactic consistent} iff
 \begin{equation*}
	\forall C_1, C_2, \mu_1, \mu_2. \quad    name(C_1) \equiv name(C_2)  \text{ and }  
	P \equiv C_1[\mu_1] \equiv C_2[\mu_2] \text{ implies }  \mu_1 \equiv \mu_2
 \end{equation*}
\end{definition}

Strong syntactic consistency adds a further constraint, by requiring also that, whenever a non-covalent bond ($\complexation{\phantom{x}}$) or a covalent bond ($\covalentbind{\phantom{x}}$) can be created between two species, then both species have the corresponding capability.
In the definition, we write $\gamma \in \mu$ as a shorthand for $\exists M.\; \mu \equiv \{ M +  \gamma \}$.

\begin{definition}[Strong Syntactic Consistency] A term $P$ is \emph{strongly syntactic consistent} iff
{
\renewcommand{\labelitemi}{$\bullet$}
\begin{itemize}
 \item $P$ is weakly syntactic consistent and
 \item $\forall C_1, C_2, \mu_1, \mu_2, N_1, N_2. $\\
    \quad$name(C_1) \equiv N_1 \text{ and } name(C_2) \equiv N_2  \text{ and }  
	P \equiv C_1[\mu_1] \equiv C_2[\mu_2]$  implies 
	\begin{align*}
		\left(  \complexation{N_2} \mu  \right)  \in \mu_1  \ \ &\Leftrightarrow\ \   \left(  \complexation{N_1} \mu  \right)  \in \mu_2 \\
		\left(  \covalentbind{N_2} \mu  \right)  \in \mu_1  \ \ &\Leftrightarrow\ \   \left(  \covalentbind{N_1} \mu  \right)  \in \mu_2
	\end{align*}
\end{itemize}
}
\end{definition}
For example, term $P_2 = \{\complexation{B}\mu_1\}.A \pipe \varnothing.B$ is weakly syntactic consistent, but not strongly syntactic consistent, since molecule $B$ does not have the capability of binding (with a non-covalent bond) to $A$. The strongly syntactic consistent term corresponding to $P_2$ is $P_3 = \{\complexation{B}\mu_1\}.A \pipe \{\complexation{A}\mu_1\}.B$.

The following proposition shows that syntactic consistency implies semantic consistency.

\begin{proposition} \emph{Syntactic Consistency} entails \emph{Semantic Consistency}, that is
\begin{align*}
  \forall P.\; P \text{ is syntactically consistent } \Rightarrow P \text{ is semantically consistent. }
\end{align*}
\end{proposition}
\begin{proof}
It is sufficient to prove that $P \ltrans{}^{*} \mu.S \pipe P'$ implies
$\exists C.\; C[\mu] \equiv P$ and $name(C) = \inactive{S}$. This proof
is done by induction on the length of the sequence of transitions $P
\ltrans{}^{*} \mu.S \pipe P'$. Let us assume that such a sequence has the
following form $P \equiv P_0 \ltrans{\alpha_1} P_1 \ltrans{\alpha_2} \cdots
\ltrans{\alpha_{n}} P_n \equiv  \mu.S \pipe P'$.\\
As regards the base case ($n = 0$), we have $P \equiv \mu.S \pipe P'$. Then,
context $C = \hole.S \pipe P'$ is such that $C[\mu] \equiv P$ and
$name(C)=\inactive{S}$.\\
As regards the induction step, let $n > 1$ and suppose that the property holds
for all $m < n$. We have $P_n \equiv \mu.S \pipe P'$ and there are two cases to
be considered: either $\mu.S$ already appeared before, i.e. $\exists
\overline{n} < n.\; P_{\overline{n}} \equiv \mu.S \pipe P''$ for some $P''$, or
not. In the first case,  by induction hypothesis, there exists a context $C$
such that $C[\mu] \equiv P$ and $name(C) = \inactive{S}$. In the
second case, $\mu.S$ has been created in the last execution step $P_{n-1}
\ltrans{\alpha_n} \mu.S \pipe P'$, where $\mu.S$ does not occur in the top-level
parallel composition, i.e. $\nexists Q.\; P_{n-1} \equiv \mu.S \pipe Q$.
According to the semantics, $P_{n-1} \ltrans{\alpha_n} \mu.S \pipe P'$ iff
$P_{n-1} \ltrans{(\emptyset,\iota)\alpha_n} \mu.S \pipe P'$ for some $\iota \in
\calN$. By rule induction on the rules \ref{eq:sem-complx}--\ref{eq:sem-parallel}
of the semantics, we prove that, for all transitions $Q
\ltrans{(\nu,\iota)\alpha_n} Q'$, for any $\mu.S$ created in the transition
there is a context $C$ such that $C[\mu] \equiv P$ and $name(C)=\inactive{S}$.
\begin{itemize}
 \item \emph{Rule \ref{eq:sem-complx}}: $\mu_1.S_1 \pipe \mu_2.S_2 
\ltrans{(\nu, \iota)\ \alpha}   \mu.( \mu_1.S_1 : \mu_2.S_2)$ with $\mu_1 \equiv
\{ X + (\nu, \iota) \complexation{\inactive{S_2}} \mu \}$\\
 By induction hypothesis, there exists a context $\overline{C}$ such that
$\overline{C}[\mu_1] \equiv P$ and $name(\overline{C}) = \inactive{S_1}$.
Therefore, context $C = \overline{C} [ \{ X + (\nu, \iota)
\complexation{\inactive{S_2}} \hole \} ]$ is such that $C[\mu] \equiv P$
and $name(C) = \inactive{( \mu_1.S_1 : \mu_2.S_2)}$.
 \item \emph{Rule \ref{eq:sem-decomplx}}:  $\mu.(\mu_1.S_1:\mu_2.S_2)
\ltrans{(\emptyset, \emptyset)\ \alpha} \mu_1.S_1 \pipe \mu_2.S_2$\\
 Let $\overline{C}$ be the context of $\mu.(\mu_1.S_1:\mu_2.S_2)$ (by induction
hypothesis). Context $\overline{C}$ must contain the portion $\hole . ( \mu_1 .
S_1 : \mu_2 . S_2 )$. Suppose that at least one of $\mu_1.S_1$ and $\mu_2.S_2$
never appeared before in any $P_m$, $m < n$ (otherwise, by induction hypothesis,
their contexts are already known). Hence, term $S = \mu.(\mu_1.S_1:\mu_2.S_2)$
has not been obtained by applying rule \ref{eq:sem-complx}, but by one of the
rules \ref{eq:sem-decomplx},\ref{eq:sem-conversion},\ref{eq:sem-losslessprod},
\ref{eq:sem-cleavagebond},\ref{eq:sem-cleavagemod}. This means that $S$
appeared literally in the inital term $P$, thus contexts $C_1$ can be obtained
from $\overline{C}$ by replacing $\hole . ( \mu_1 . S_1 : \mu_2 . S_2 )$ with
$\mu . ( \hole . S_1 : \mu_2 . S_2 )$ and contexts $C_2$ can be obtained
from $\overline{C}$ by replacing $\hole . ( \mu_1 . S_1 : \mu_2 . S_2 )$ with
$\mu . ( \mu_1 . S_1 : \hole . S_2 )$. Contexts $C_1$ and $C_2$ are such that
$C_1[\mu_1] \equiv P$ with $name(C_1) = \inactive{S_1}$ and $C_2[\mu_2] \equiv
P$ with $name(C_2) = \inactive{S_2}$.
 \item \emph{Rule \ref{eq:sem-conversion}}: $\mu_1.S_1 \ltrans{(\nu, \iota)\
\alpha} \mu.S \pipe P'$ with  $\mu_1 = \{ X + (\nu, \iota) \conversion (\mu.S
\pipe P') \}$\\
 Let $\overline{C}$ be the context of $\mu_1.S_1$. Then the context for $\mu.S$
is $C = \overline{C} [ \{ X + (\nu, \iota) \conversion (\hole.S \pipe
P')\} ]$.
 \item \emph{Rule \ref{eq:sem-losslessprod}}: analogous to rule \ref{eq:sem-conversion}.
 \item \emph{Rule \ref{eq:sem-covalentbind}}: analogous to rule \ref{eq:sem-complx}.
 \item \emph{Rule \ref{eq:sem-cleavagebond}}: $\mu_1.S_1 \pipe \mu'.\bond{(\mu_2.S_2) (\mu_3.S_3)} \ltrans{(\nu, \iota)\ \alpha} \mu_1.S_1 \pipe \mu_2.S_2 \pipe \mu_3.S_3$\\
 Similarly to rule \ref{eq:sem-decomplx}, if either $\mu_2.S_2$ or $\mu_3.S_3$
did not appear before, their contexts can be obtained from context
$\overline{C}$ of $\mu'.\bond{(\mu_2.S_2) (\mu_3.S_3)}$. Context $\overline{C}$
must contain $\hole.\bond{(\mu_2.S_2) (\mu_3.S_3)}$. We obtain context $C$
by replacing $\hole.\bond{(\mu_2.S_2) (\mu_3.S_3)}$ with $\mu'.\bond{(\hole.S_2)
(\mu_3.S_3)})$ in $\overline{C}$ for $\mu_2.S_2$. Similarly, we obtain context
$C$ by replacing $\hole.\bond{(\mu_2.S_2) (\mu_3.S_3)}$ with
$\mu'.\bond{(\mu_2.S_2) (\hole.S_3)})$ in $\overline{C}$ for $\mu_3.S_3$.
 \item \emph{Rule \ref{eq:sem-covalentmod}}: $\mu_1.S_1  \ltrans{(\nu, \iota)\
\alpha}   \mu.\bond{q(\mu_1.S_1)}$ with $\mu_1 = \{ X + (\nu, \iota)
\covalentmod{q} \mu \}$\\
 Let $\overline{C}$ be the context of $\mu_1.S_1$. Then the context for
$\mu.\bond{q(\mu_1.S_1)}$ is $C = \overline{C} [\{ X + (\nu, \iota)
\covalentmod{q} \hole \}]$.
  \item \emph{Rule \ref{eq:sem-cleavagemod}}:  analogous to rules \ref{eq:sem-decomplx} and \ref{eq:sem-cleavagebond}.
 \item \emph{Rule \ref{eq:sem-parallel}}: $P \pipe Q \ltrans{(\nu', \iota)\ \alpha} P' \pipe Q$\\
 Since $Q$ is not modified by the transition, only terms $\mu.S$ in $P'$ could have been created by the transition. The contexts of any $\mu.S$ in $P'$ is given by the induction hypothesis on the rule. \qedhere
\end{itemize}

\end{proof}

\section{An Example of Modeling}
In this section we show an example of a real molecular interaction map, taken from \cite{K99}, and we show the corresponding term in MIM calculus.
Differently from Kohn maps, the MIM calculus can contain multiple molecules of a same molecular species, thus it can describe the evolution of the system starting from an initial configuration.

\begin{figure}[htp]
 \centering
 \includegraphics[scale=0.25]{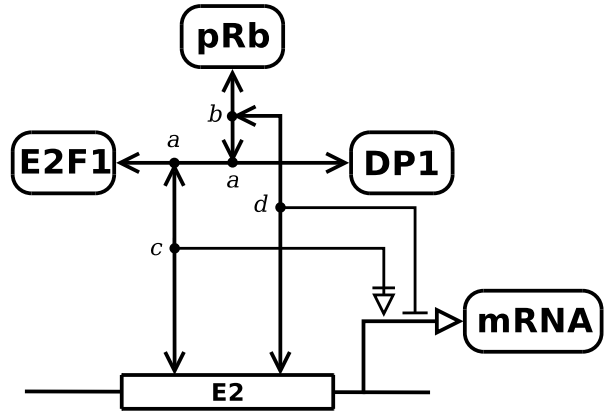}
 \caption{Molecular interaction map representing interactions among E2F1, DP1, and pRb.}
 \label{fig:kohnmap}
\end{figure}

The example in \cite{K99} presents a comprehensive molecular interaction map of regulators of cell cycle and DNA repair processes. The presented map is limited to the events in the mammalian cell nucleus. We consider here only the interaction between a protein of the E2F family and a gene promoter E2. This interaction is an important part of the cell cycle. The transcription of the gene is activated or inhibited by the binding of different complexes with the promoter. The molecular interaction map representing the interactions among stimulatory and inhibitory complexes of E2F1, DP1 and pRb is shown in Figure \ref{fig:kohnmap}.

The E2F1:DP1 dimer (indicated by node (a) in Figure \ref{fig:kohnmap}) and the (E2F1:DP1):pRb trimer (node (b)) can be bound to the promoter element E2. When the E2F1:DP1 dimer is bound to E2 the transcription activity is stimulated, while when the (E2F1:DP1):pRb trimer is bound to E2 the transcription is inhibited. We represent each species involved by a MIMc term with the capabilities of the species itself. In particular we use particular terms for representing the promoter E2 and the DNA. In the Kohn map the DNA is implicitly represented, but in a MIMc term DNA must be represented explicitly, for assigning to it the capability of producing the mRNA. Remark that we can have multiple copies of the species E2F1, DP1, and pRb in a MIMc term representing the system. Coherently with the cell system we can have only one copy of the DNA and of the gene promoter E2.
Each basic element is identified by an elementary species name $E2F1,DP1,pRb,E2,DNA,mRNA \in \calE$.

The E2F1 species can be represented by the following term:
\begin{align*}
 &\{ \complexation{DP1} \{ \complexation{E2} \varnothing  +  \complexation{pRb} \{ \complexation{E2} \varnothing \}    \}  \}.E2F1  = \mu_1.E2F1
\end{align*}
which states that $E2F1$ can be bound to $DP1$, and then the dimer can be bound to $E2$. Note that the stimulation of DNA transcription by the trimer is not modeled among the capability of the species, which are just empty. Instead, this behavior is captured by the DNA process, shown in the following. Moreover, the $E2F1:DP1$ dimer can be bound to $pRb$ and then to promoter $E2$ to inhibit the transcription. As for stimulation, inhibition is captured inside the definiton of the DNA. Species $DP1$ and $pRb$ are represented by the terms:
\begin{align*}
 &\{ \complexation{E2F1} \{ \complexation{E2} \varnothing  +  \complexation{pRb} \{ \complexation{E2} \varnothing \}    \}  \}.DP1 = \mu_2.DP1 \\
 &\{ \complexation{E2F1:DP1} \{ \complexation{E2} \varnothing  \} \}.pRb = \mu_3.pRb
\end{align*}
Finally, the promoter E2 and the DNA can be represented by the terms:
\begin{align*}
 &\{  \complexation{E2F1:DP1} \varnothing     +      \complexation{(E2F1:DP1):pRb} \varnothing   \}.E2 = \mu_4.E2 \\
 &\{  (\nu_{DNA},\iota_{DNA})   \losslessprod  mRNA \}.DNA \quad \text{ with } \left\lbrace 
     \begin{array}{l}
      \nu_{DNA} = \{ (E2F1:DP1):E2\}\\
      \iota_{DNA} = \{((E2F1:DP1):pRb):E2\}
     \end{array}
  \right.
\end{align*}
The lossless production of $mRNA$ by the $DNA$ is regulated by the presence/absence of the two complexes $(E2F1:DP1):E2$ and $((E2F1:DP1):pRb):E2$. In particular, the former complex represents a promoter (triggering the reaction), while the latter represents an inhibitor for the reaction.

An initial configuration in which two molecules of species E2F1, DP1 and pRb are present is represented by the following MIMc term
\begin{align*}
 P_1 &= \mu_1.E2F1 \pipe  \mu_1.E2F1 \pipe
 \mu_2.DP1   \pipe  \mu_2.DP1  \pipe
 \mu_3.pRb   \pipe  \mu_3.pRb  \pipe
 \mu_4.E2 \\
 &\pipe \{  (\nu_{DNA},\iota_{DNA})   \losslessprod  mRNA \}.DNA 
\end{align*}
The term can evolve towards different configurations. For example, after a complexation between $E2F1$ and $DP1$ occurs, the processes $\mu_1.E2F1$ and $\mu_2.DP1$ are replaced by the following term, representing a complex with name $E2F1:DP1$:
\begin{align*}
\{ \complexation{E2} \varnothing  +  \complexation{pRb} \{ \complexation{E2} \varnothing \}    \}. \left(
 \mu_1.E2F1 : \mu_2.DP1  \right)
\end{align*}
Thus the whole term becomes:
\begin{align*}
 P_2 &=  \mu_1.E2F1 \pipe
  \mu_2.DP1  \pipe
 \mu_3.pRb   \pipe  \mu_3.pRb  \pipe
 \mu_4.E2 \pipe \{  (\nu_{DNA},\iota_{DNA})   \losslessprod  mRNA \}.DNA  \\
 &\pipe  \{ \complexation{E2} \varnothing  +  \complexation{pRb} \{ \complexation{E2} \varnothing \}    \}. \left(
 \mu_1.E2F1 : \mu_2.DP1  \right) 
\end{align*}
As a further evolution step we may have the binding of the dimer $E2F1:DP1$ to the promoter $E2$. The resulting term is:
\begin{align*}
 P_3 &=  \mu_1.E2F1 \pipe
  \mu_2.DP1  \pipe
 \mu_3.pRb   \pipe  \mu_3.pRb  \pipe
  \{  (\nu_{DNA},\iota_{DNA})   \losslessprod  mRNA \}.DNA  \\
 &\pipe  
 \varnothing. \left(
 (\{ \complexation{E2} \varnothing  +  \complexation{pRb} \{ \complexation{E2} \varnothing \}    \}. ( \mu_1.E2F1 : \mu_2.DP1  ))  :  \mu_4.E2   \right) 
\end{align*}

As an example of derivation, we show how the semantics is applied to the term $P_1$ above obtaining the term $P_2$ in a single reduction step.
For the sake of readability, we write the terms $P_1$, $P_2$ as:
\begin{align*}
P_1 &= \mu_1.E2F1 \pipe \mu_2.DP1 \pipe R \\
P_2 &= \mu. \left( \mu_1.E2F1 : \mu_2.DP1 \right) \pipe R
\end{align*}
where 
\begin{align*}
R &= \mu_1.E2F1 \pipe \mu_2.DP1  \pipe \mu_3.pRb   \pipe  \mu_3.pRb  \pipe \mu_4.E2 \pipe \{ (\nu_{DNA},\iota_{DNA})   \losslessprod  mRNA \}.DNA \\
\mu &= \{ \complexation{E2} \varnothing + \complexation{pRb} \{ \complexation{E2} \varnothing \} \} .
\end{align*}
The transition $P_1 \ltrans{(\nu', \iota)} P_2$, with $\nu' = \iota = \varnothing$, is obtained with the following derivation, by using the rules of the semantics:
\begin{gather*}
\InfRule{
 	\InfRule{\mu_1 = \{ X + (\nu, \iota) \complexation{DP1} \mu \}
 		\AND \nu = \iota = \varnothing
 		\AND  E2F1,DP1 \not\in \iota 
 	}
	{ 
 		\mu_1.E2F1 \pipe \mu_2.DP1 \ltrans{(\nu, \iota)\ \ E2F1 \vcomplexation DP1} \mu. \left( \mu_1.E2F1 : \mu_2.DP1 \right) 
 	} 
 	\quad
 	\stacked{\phantom{XXXX}}{\inactive{R} \cap \iota = \varnothing   \AND  \nu' = \nu \setminus \inactive{R} = \varnothing}
}
{
	\mu_1.E2F1 \pipe \mu_2.DP1 \pipe R
	\ltrans{(\nu', \iota)\ \ E2F1 \vcomplexation DP1}
	\mu. \left( \mu_1.E2F1 : \mu_2.DP1 \right) \pipe R 
}%
\end{gather*}
Finally, according to property (\ref{eq:sem0}), we have the transition $P_1 \ltrans{E2F1 \vcomplexation DP1} P_2$.

\section{Conclusions}
Using formal methods for studying biological systems is an interesting approach that allows using many different analiysis techniques.
We have defined the MIM calculus, a new calculus with high-level operators directly inspired by Molecular Interaction Maps (MIM), a graphical notation used in biology. 
This approach allows exploiting the features of process calculi, such as incremental definition of models, techniques for analysis and verification of properties, and the development of simulators. Moreover, the correspondence of the operators of the calculus with biological reactions allows an easy translation of Molecular Interaction Maps into terms of the MIM calculus.
We have studied conditions under which a term of the MIM calculus is a formal representation of a MIM diagram, and we have provided different consistency definitions for the terms of the MIM calculus.

In the future, we plan to investigate the properties of calculus, such as its expressiveness, and to develop different extensions for a better description of biological systems.

\bibliographystyle{eptcs}

\bibliography{biblio}{}

\end{document}

%% file: macros/generic.tex
\newcommand{\IS}{\mathrel{\makebox[2em]{::=}}}
\newcommand{\OR}{\quad\big|\quad}

\newcommand{\InfRule}[2]{\dfrac{\; #1 \;}{\; #2 \;}}

\newcommand{\AND}{\qquad}
\newcommand{\NR}{\\[\smallskipamount]}
\newcommand{\stacked}[2]{{\displaystyle{#1 \atop #2}}}

\newcommand{\ltrans}[1]{\xrightarrow{#1}}

\newcommand{\hole}{\square}

\usepackage{mathrsfs}
\DeclareMathOperator{\PowerSet}{\mathbf{P}}

\newcommand{\calC}{\mathcal{C}}

\newcommand{\calE}{\mathcal{E}}

\newcommand{\calN}{\mathcal{N}}

\newcommand{\calP}{\mathcal{P}}

\newcommand{\calS}{\mathcal{S}}

\newcommand{\calX}{\mathcal{X}}

%% file: macros/mimcalc.tex
\newcommand{\inactive}[1]{\left\lfloor #1\right\rfloor}
\newcommand{\bond}[1]{\overline{#1}}
\newcommand{\nil}{\mathbf{0}}

\newcommand{\pipe}{\mid}

\usepackage{extarrows}
\usepackage{morextarrows}
\usepackage{mimsymbols}

\newcommand{\complexation}[1]{\mathord{\xlongrightarrow{#1}}}

\newcommand{\conversion}{\mathord{\xrightblackarrow{\phantom{xx}}}}
\newcommand{\covalentbind}[1]{\mathord{\xlongequal{#1}}}
\newcommand{\covalentmod}[1]{\mathord{\stackrel{#1}{\Longrightarrow}}}

\newcommand{\cleavage}[1]{\mathord{\stackrel{#1}{\text{\large\mylightning}}}}
\newcommand{\losslessprod}{\mathord{\xrightwhitearrow{\phantom{xx}}}}

\newcommand{\vcomplexation}{\mathord{\leftrightarrow}}
\newcommand{\vdecomplexation}{\mathord{\nleftrightarrow}}
\newcommand{\vconversion}{\conversion}
\newcommand{\vlosslessprod}{\losslessprod}
\newcommand{\vcovalentbind}{\covalentbind{\,}}
\newcommand{\vcleavage}{\cleavage{}}
\newcommand{\vcovalentmod}{\mathord{\,\Rightarrow}}

%% file: macros/thenvs.tex
\theoremstyle{plain}
\newtheorem{theorem}{Theorem}[section]

\newtheorem{proposition}[theorem]{Proposition}

\theoremstyle{definition}
\newtheorem{definition}{Definition}[section]

\theoremstyle{remark}